\documentclass{PoS}

\usepackage{graphicx}
\usepackage{natbib}

\usepackage{amsmath}
\usepackage{amssymb}
\usepackage{threeparttable}
\usepackage{epsfig}

\title{Pulse-to-pulse variability of bright accreting pulsars}

\ShortTitle{Pulse-to-pulse variability}

\author{\speaker{Dmitry Klochkov}\\
        Institut f\"ur Astronomie und Astrophysik, T\"ubingen, Germany\\
        E-mail: \email{klochkov@astro.uni-tuebingen.de}}

\author{Andrea Santangelo\\
        Institut f\"ur Astronomie und Astrophysik, T\"ubingen, Germany\\
        E-mail: \email{santa@astro.uni-tuebingen.de}}

\author{R\"udiger Staubert\\
        Institut f\"ur Astronomie und Astrophysik, T\"ubingen, Germany\\
        E-mail: \email{staubert@astro.uni-tuebingen.de}}

\author{Richard E. Rothschild\\
        Center for Astrophysics and Space Sciences, San Diego, USA\\
        E-mail: \email{rrothschild@ucsd.edu}}

\abstract{
In addition to coherent pulsation, many accreting neutron stars exhibit
flaring activity and strong aperiodic variability on time scales
similar to or shorter than their pulsation period. Such a behavior
shows that the accretion flow in the vicinity of the accretor must be
highly non-stationary. Although from the theoretical point of view
the problem of non-stationary accretion has been addressed by many
authors, observational study of this phenomenon is often problematic
as it requires very high statistics of X-ray data and a specific analysis
technique. In our research we used high-resolution
data taken with \textsl{RXTE} and \textsl{INTEGRAL} 
on a sample of bright transient and persistent
pulsars, to perform an in-depth study of their variability on time scales
comparable to the pulsation period -- ''pulse-to-pulse
variability''. The high-quality data allowed us to collect individual
pulses of different amplitude and reveal differences in their spectra
(such an analysis we refer to as ''pulse-to-pulse spectroscopy''). The described
approach allowed us for the first time to study luminosity-dependence
of pulsars' X-ray spectra in observations where the
averaged (over many pulse cycles) luminosity of the source remained constant
and discuss them in the frame of the current physical models of
the accretion flow close to the neutron star surface.
}

\FullConference{8th INTEGRAL Workshop ''The Restless Gamma-ray Universe''- Integral2010,\\
		September 27-30, 2010\\
		Dublin Ireland}

\begin{document}

\section{Introduction}

Strong aperiodic variability is a well known 
characteristic of many X-ray binary systems
-- both black hole candidates and accreting neutron stars.
%First discovered in
%black hole candidates (with Cyg~X-1 being the most remarkable case) 
%it was later shown to be a common feature also among accreting neutron stars
%\citep[see e.g.][]{Belloni:Hasinger:90}. 
Although \emph{homogeneity}
and \emph{stationarity} of the accretion flow is often assumed in calculation
dealing with the structure of the X-ray emitting region above neutron star,
%(as it greatly simplifies the mathematical treatment of the problem),
the observed variability clearly indicates that the accretion flow in 
the vicinity of the accretor must be highly non-stationary. 
%From the theoretical point of view the problem of non-stationary accretion 
%has been addressed e.g. by 
\citet{Morfill:etal:84,Demmel:etal:90,
Orlandini:Boldt:93} have shown that the inhomogeneity
of the flow is generally expected in accreting pulsars and that it does not 
necessarily arise from the original inhomogeneity of the accreted matter 
%(e.g. ''lumps'' in the stellar wind or variations of matter supply from
%the donor star) 
but is rather produced by instabilities close to the 
magnetospheric boundary of the neutron star.

So far, the aperiodic variability of accreting pulsars has mostly been 
studied by means of power spectra of high-time-resolution light curves 
\citep[e.g.][]{Belloni:Hasinger:90,Revnivtsev:etal:09}. In our work, however,
we focus on the variable profile shape of \emph{individual} X-ray pulses
exhibited by many accreting neutron stars -- \emph{pulse-to-pulse variability}.
Such a variability seems to be a common phenomenon among accreting pulsars
and has been reported by several authors 
%for bright outbursts of transient sources 
(\citealt{Frontera:etal:85} for A0535+26, \citealt{Kretschmar:etal:00}
for Vela\,X-1, \citealt{Tsygankov:etal:07} for 4U\,0115+63). However, a
detailed study of pulse-to-pulse variability is usually limited by
the photon statistics.
%, especially for relatively fast pulsars (with the
%pulse period of a few second or less). 
In the present work, 
we used the archival data from high-sensitivity X-ray detectors onboard 
\textsl{RXTE} and \textsl{INTEGRAL} taken on a sample of bright accreting pulsars.
%V0332+53, 4U\,0115+63, A\,0535+26, and Her\,X-1. 
Using a special analysis
technique, we collected individual pulses of different amplitude
and studied differences in their X-ray spectra.

%%%%%%%% Table 1 %%%%%%%%
\begin{table}
  \centering
  \caption{Observations used for the pulse-to-pulse analysis}
  \label{obs}
  \begin{threeparttable}
    %\vspace*{1mm}
    \begin{tabular}{l l l l l}
      \hline\hline
      Source name & Instrument   & middle MJD & Exposure (ksec) & Remarks \\
      \hline
      V0332+53    & \textsl{RXTE}     & 53354 & 23.7 &Giant outburst 2004 \\
      4U\,0115+63 & \textsl{RXTE}     & 51249 & 32.8 & Giant outburst 1999\\
      A\,0535+26  & \textsl{RXTE/Integral}  & 53615 & 30.8/6.6 & Normal outburst 2005\\
       %           & \textsl{INTEGRAL} & 53615          & 6.6  & Normal outburst 2005\\
      Her X-1     & \textsl{RXTE}     & 52600 & 98.7 & Main-On state\\
      \hline
    \end{tabular}
  \end{threeparttable}
\end{table}
%%%%%%%%%%%%%%%%%%%%%%%%%

%-------------------------------------------------------------------
\section{Observational data and analysis technique}

%%%%%%%%%%%%%% Figure 1 %%%%%%%%%%%%
\begin{figure}
\centering
\begin{minipage}{0.65\textwidth}
\epsfig{file=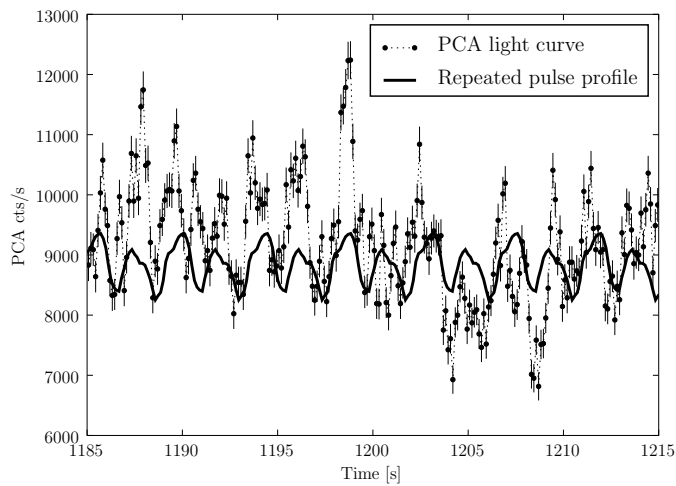, width=1.0\textwidth}
\end{minipage}
\begin{minipage}{0.33\textwidth}
\caption{Sample light curve of V0332+53 during its 2005 giant outburst obtained
      with \textsl{RXTE/PCA} (summed over all channels). The solid curve
      shows the repeated pulse profile obtained by folding of a longer data
      sample. Dramatic pulse-to-pulse variability is clearly seen.
}
\label{lc_p2p}
\end{minipage}
\end{figure}
%%%%%%%%%%%%%%%%%%%%%%%%%%%%%%%%%%%%

For our analysis we used the data taken with the \textit{RXTE} satellite
during intense outbursts of the transient high mass X-ray binaries 
V0332+53, 4U\,0115+63, and A\,0535+26 and during a \emph{main-on} state of the
persistent intermediate mass X-ray binary Her~X-1 (such states of
high X-ray flux repeat in the system about every 35\,days).
%most probably reflecting periodic obscuration of the source by a precessing tilted
%accretion disk, see e.g. \citealt{Klochkov:etal:06} and references
%therein). 
For A\,0535+26 we also used the data taken with \textsl{INTEGRAL}
simultaneously with the \textsl{RXTE} observations.

For each of the three transient sources we used a short continuous set 
of pointings covering $\sim$1 day or less of the bright part of an
outburst so that the average flux level 
was not changing significantly within the observations.
For the persistent pulsar Her~X-1 we used the data from the
main-on state of the source 
%corresponding to the 35\,d cycle No.\,323 
%(according to the numbering convention adopted in \citealt{Staubert:etal:83})
which is best covered by \textsl{RXTE} observations
(we restricted ourselves with the data from the middle part of the main-on
where the flux does not change significantly).
%, i.e. where the obscuration
%by the accretion flow is minimal. 
%In this case the data are spread over 
%$\sim$5\,d. 
The observational data are summarized in Table\,\ref{obs} 

For the pulse-to-pulse analysis we extracted 
a high-resolution light curve of each source where single pulses are 
clearly distinguishable. An example of such a light curve obtained with 
\textsl{RXTE/PCA} on V0332+53 is shown in Fig.\,\ref{lc_p2p} where
strong variability of the profile shape from one pulse to the next
is visible. 
%Then we selected a pulse-phase interval
%covering the brightest peak of the source's pulse profile 
%(such intervals will be referred to as \emph{pulses}).
For each individual pulsation cycle in the extracted light curve
we measured the average count rate within a pre-selected pulse-phase
interval covering the brightest peak of the source's pulse profile
(the \emph{pulse}), 
which we call \emph{amplitude} of a pulse. 
As can be seen in Fig.\,\ref{lc_p2p} the amplitudes of individual pulses
vary over a broad range. This allows one to explore the variation of
the source X-ray spectrum as a function of the pulse amplitude, which is the
central point of the presented analysis. 
%To perform such a study
We collected the data from pulses within narrow ranges of amplitude 
-- amplitude bins -- and for each bin performed spectral extraction.
As a result, we obtained X-ray spectra as a function
of the individual pulse amplitude or pulse flux. 

%-------------------------------------------------------------------
\section{Results}

For each pulsar from our sample we obtained a series of broad band
($\sim$3--80\,keV) X-ray spectra corresponding to different
pulse fluxes. The spectral continuum was modeled using a usual 
powerlaw-cutoff function. The four accreting pulsars are well established 
cyclotron line (or \emph{Cyclotron Resonant Scattering Feature, CRSF})
sources. 
%At least three of them exhibit more than one 
%CRSF harmonic. 
The absorption features 
%that are clearly seen in our
%pulse-flux resolved spectra
were modeled using a multiplicative Gaussian absorption line model.
%with a Gaussian
%optical depth profile. 
We checked our fits for intrinsic dependencies between
different pairs of the model parameters using contour plots
to exclude any degeneracies. The presented
results were also found to be stable with respect to the choice of
different spectral functions (e.g. \texttt{cutoffpl} vs. 
\texttt{powerlaw*hichecut} XSPEC models). We are, therefore, confident
that the variability reported below arises from the data and reflects
real physics.

%- - - - - - - - - - - - - - - - - - - - - - - - - - - - - - - - - - 
%\subsection{V0332+53}

\begin{figure}
\centering
\begin{minipage}{0.45\textwidth}
  \epsfig{file=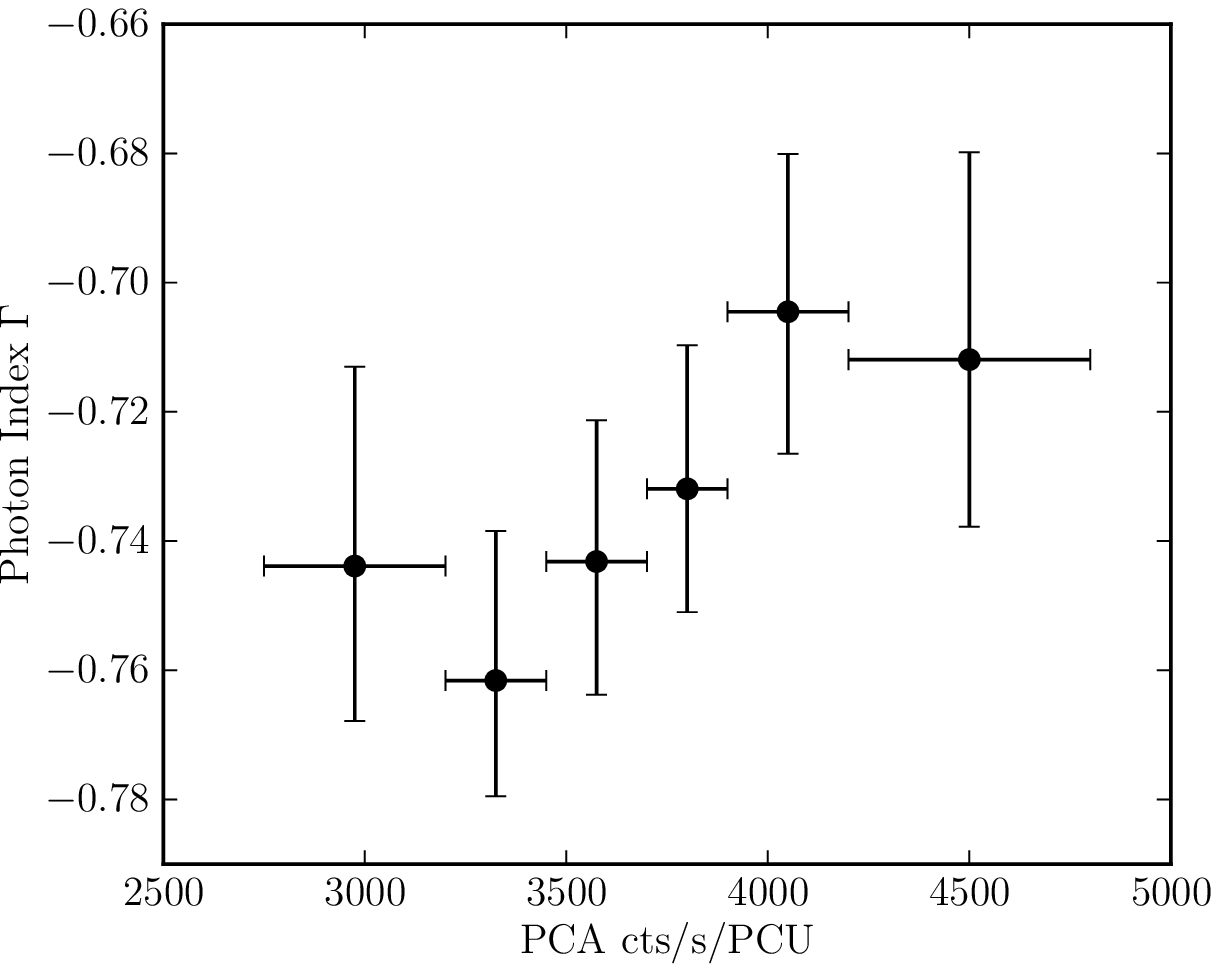, width=\textwidth}
\end{minipage}
\begin{minipage}{0.45\textwidth}
  \epsfig{file=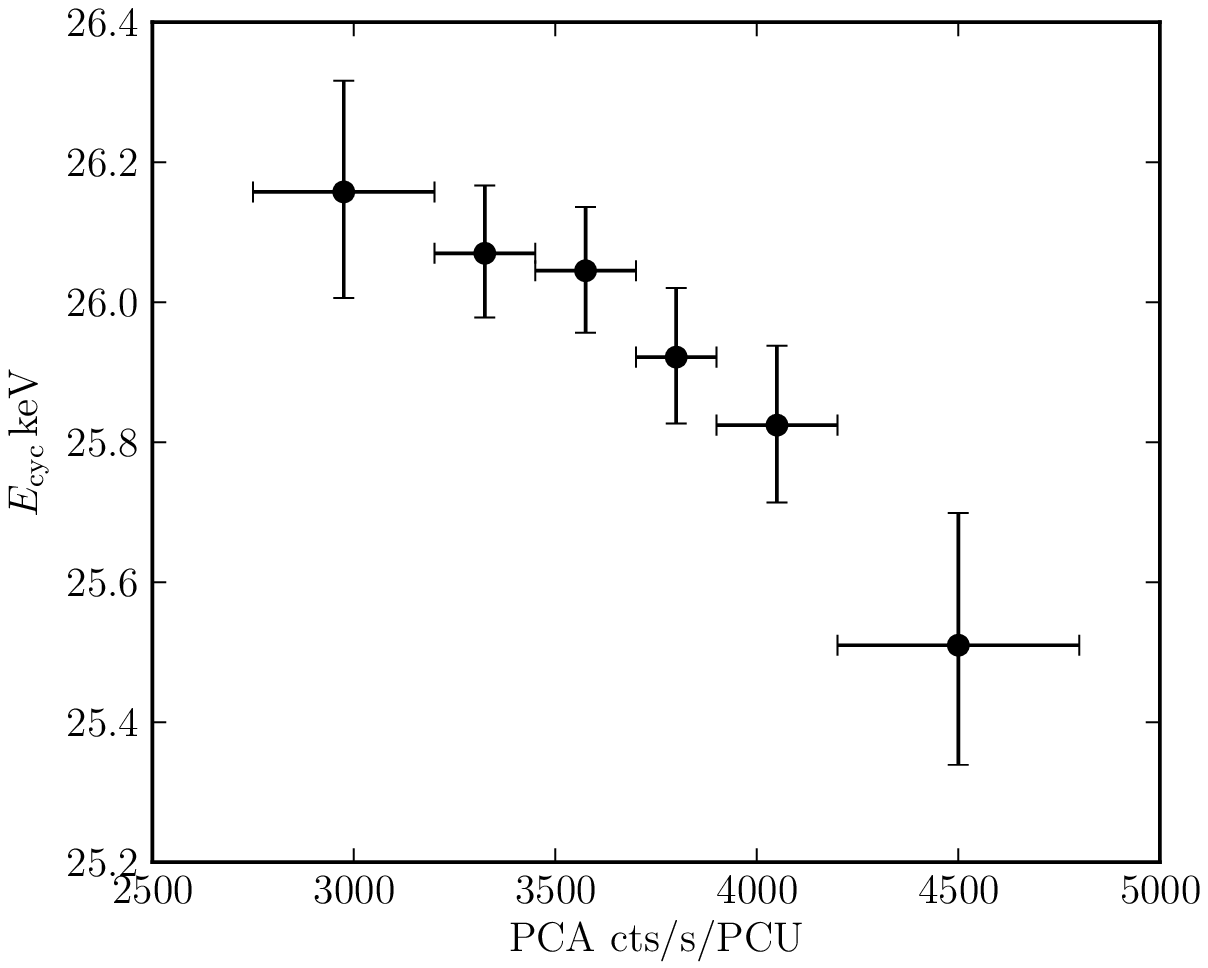, width=\textwidth}
\end{minipage}
\caption{The variation with the pulse amplitude of the photon index
(left) and the fundamental cyclotron line centroid energy (right) measured
for V0332+53 using the \textsl{RXTE} data. The vertical bars
in all plots indicate uncertainties at the 90\% c.l.}
\label{v0332frs}
\end{figure}

\textbf{V0332+53}.
%The dynamical range of the pulse amplitude variations reaches
%a factor of $\sim$$1.5$. 
In this source we were able to measure 
the changing of two spectral parameters: the photon
index $\Gamma$ and the fundamental cyclotron line energy $E_{\rm cyc}$
as shown in Fig.\,\ref{v0332frs}. There is an indication of softening of 
the spectral continuum with increasing pulse flux (see the left panel).
The negative correlation of the cyclotron line energy with the pulse amplitude
clearly appears on the right panel. A negative correlation of 
$E_{\rm cyc}$ with the \emph{averaged} flux level has been reported for this
source by \citet{Mowlavi:etal:06,Tsygankov:etal:10} during the rise and
decay of strong outbursts. In our analysis, however, we observe the
correlation during a short set of pointing 
where the averaged flux level stays constant.

%%%%%%%%%%%%%% Figure 2 %%%%%%%%%%%%
%%%%%%%%%%%%%%%%%%%%%%%%%%%%%%%%%%%%

%- - - - - - - - - - - - - - - - - - - - - - - - - - - - - - - - - - 
%\subsection{4U\,0115+63}

\begin{figure}
\centering
\begin{minipage}{0.32\textwidth}
  \epsfig{file=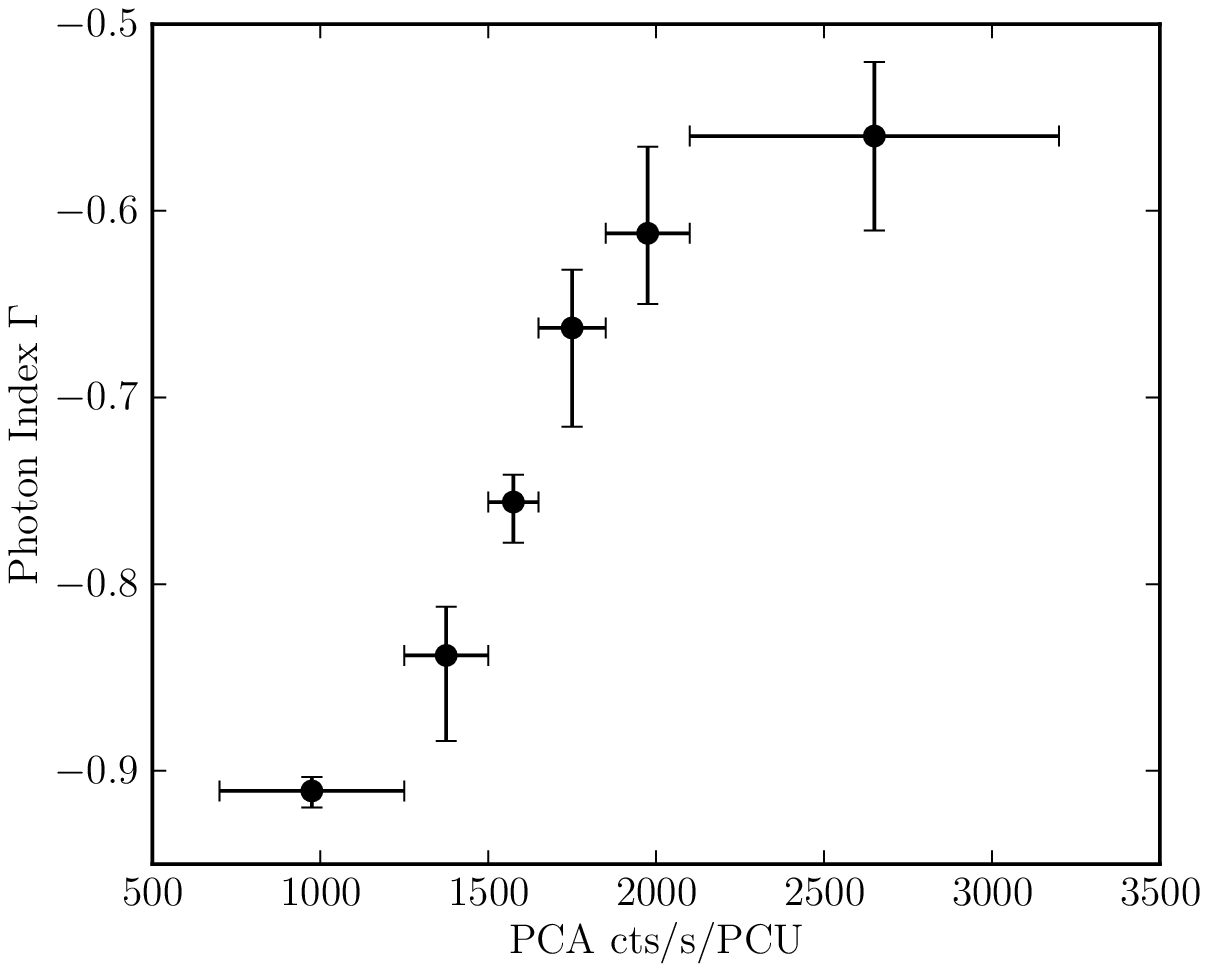, width=1.12\textwidth}
\end{minipage}
\begin{minipage}{0.32\textwidth}
  \epsfig{file=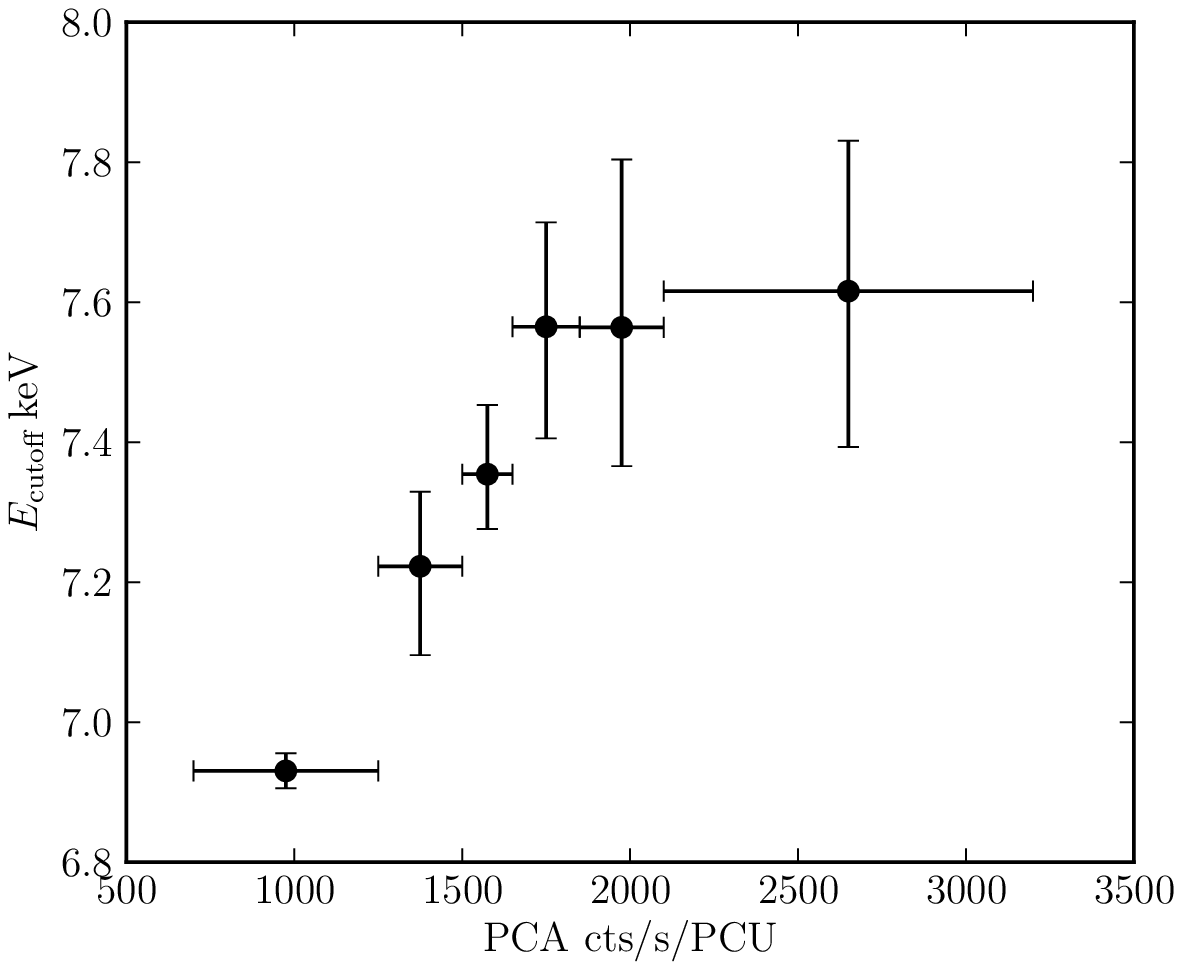, width=1.12\textwidth}
\end{minipage}
\begin{minipage}{0.32\textwidth}
  \epsfig{file=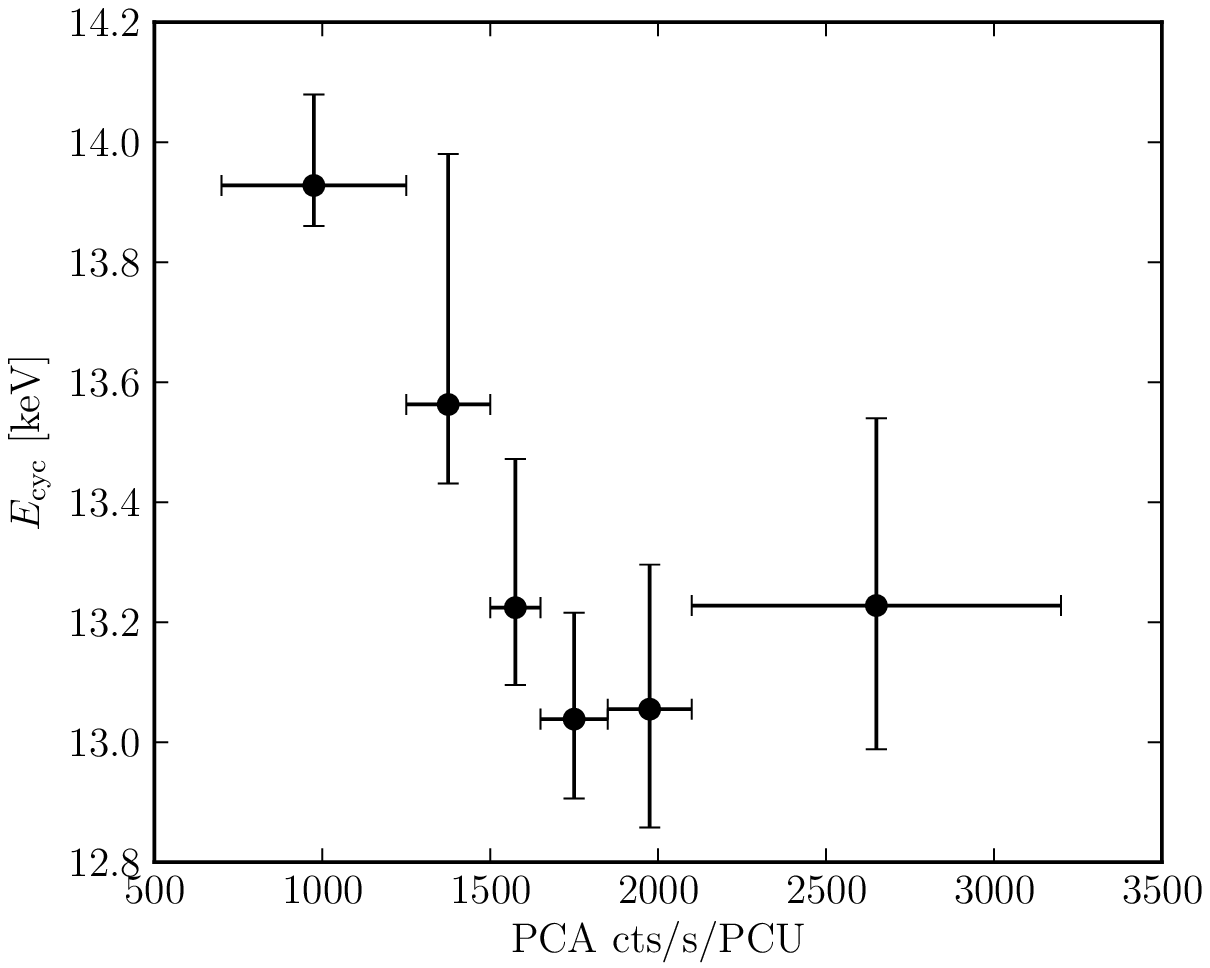, width=1.12\textwidth}
\end{minipage}
\caption{The variation with the pulse amplitude of the photon index
(left), the exponential cutoff energy (middle), and the fundamental 
cyclotron line centroid energy (right) measured
for 4U~0115+63 using the \textsl{RXTE} data. 
%The vertical bars indicate uncertainties at the 90\% c.l.
}
\label{x0115frs}
\end{figure}

\textbf{4U\,0115+63.}
%The amplitude of variations in the pulse fluxes exhibited by 4U\,0115+63 
%in our data sample reached a factor of $\sim$$2.5$ allowing somewhat deeper
%study of continuum changing with pulse flux compared to the previous source.
The measured dependencies of the photon index $\Gamma$, exponential
cutoff energy $E_{\rm cutoff}$, and the cyclotron line energy $E_{\rm cyc}$
are presented in Fig.\,\ref{x0115frs}. The positive correlation of 
$\Gamma$ with pulse amplitude (left panel) is similar to that measured in
V0332+53. The cutoff energy $E_{\rm cutoff}$ 
%(which was not found to vary
%significantly in V0332+53) in this source
is strongly positively correlated with pulse
flux (middle panel). A strong decrease of the cyclotron
line energy with pulse flux is seen in the right panel. Similar to 
V0332+53, a negative correlation of $E_{\rm cyc}$ with the averages 
(over many pulsation cycles) flux was reported during gradual flux
changes in giant outbursts of this source 
\citep{Mihara:etal:98,Tsygankov:etal:07}.
Again, in our work we find a similar correlation on the bases of
pulse-to-pulse variation while the averaged flux level does not change
significantly. 
%during the observations.

%%%%%%%%%%%%%% Figure 3 %%%%%%%%%%%%
%%%%%%%%%%%%%%%%%%%%%%%%%%%%%%%%%%%%

%- - - - - - - - - - - - - - - - - - - - - - - - - - - - - - - - - - 
%\subsection{Her X-1}

\begin{figure}
\centering
\begin{minipage}{0.45\textwidth}
  \epsfig{file=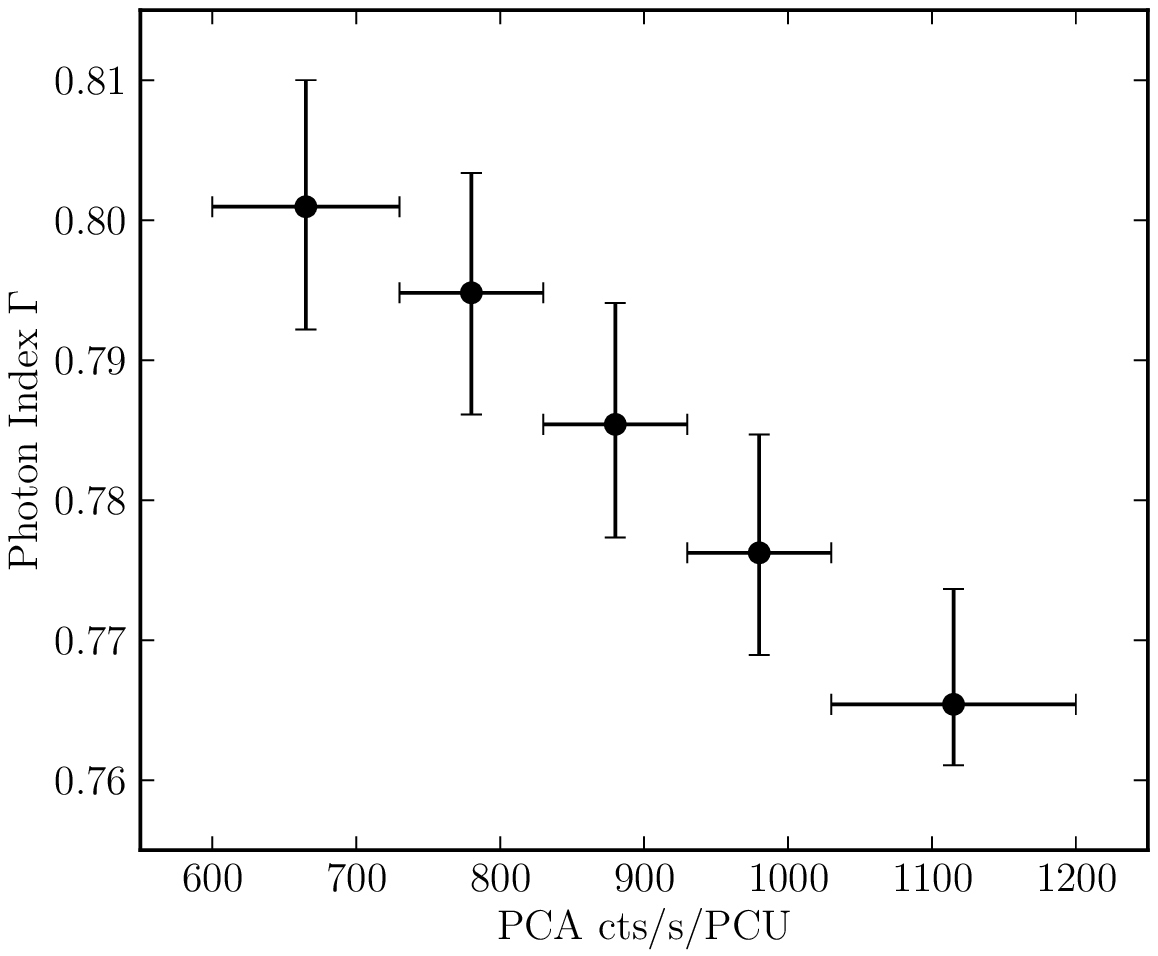, width=\textwidth}
\end{minipage}
\begin{minipage}{0.45\textwidth}
  \epsfig{file=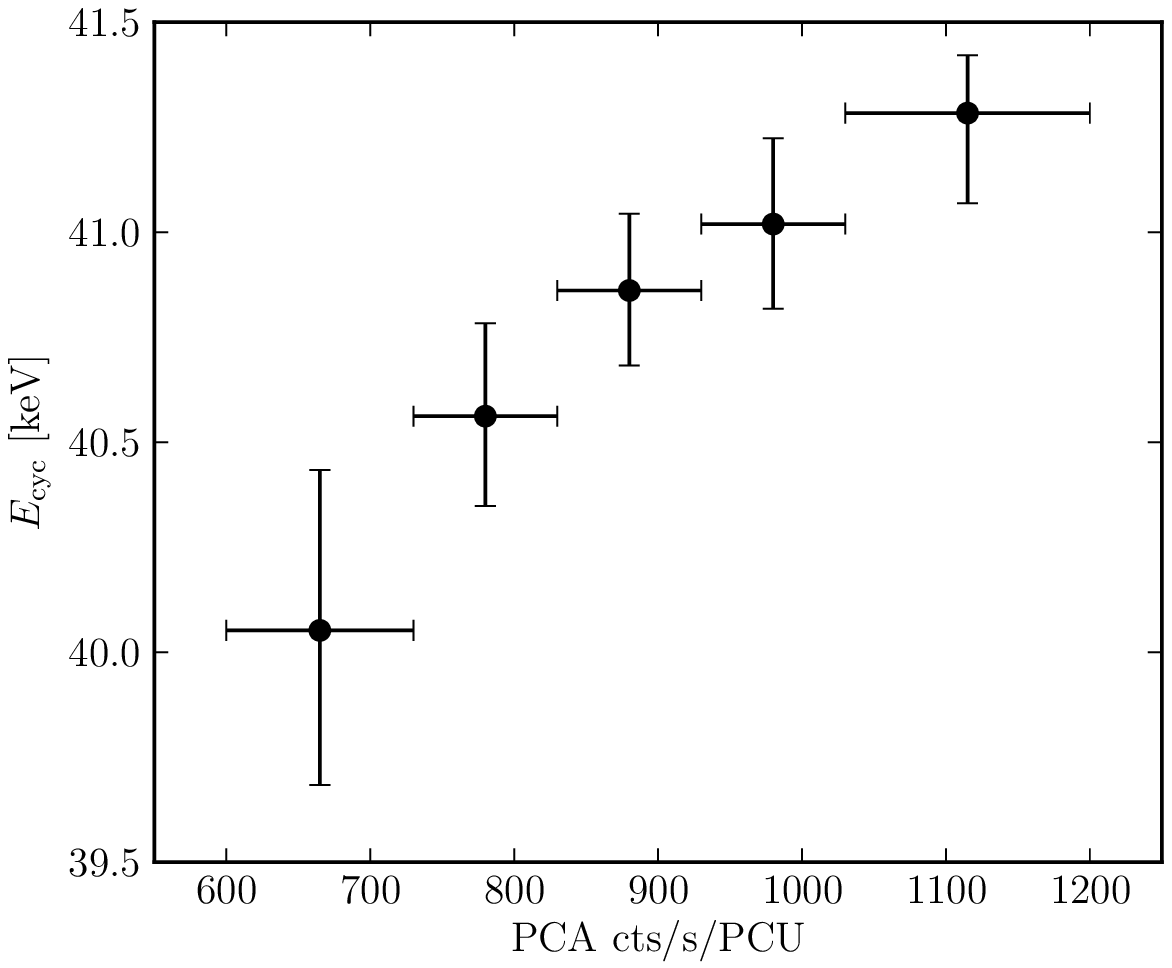, width=\textwidth}
\end{minipage}
\caption{The variation with the pulse amplitude of the photon index
(left) and the cyclotron line centroid energy (right) measured
in Her X-1 using the \textsl{RXTE} data from the middle of a singe main-on
state. 
%The vertical bars indicate uncertainties at the 90\% c.l.
}
\label{herx1frs}
\end{figure}

\textbf {Her X-1}. This source is particularly interesting as far as the
flux-dependence of the X-ray spectrum is concerned. First, it is a
persistent pulsar showing only minor variation of the averaged 
intrinsic X-ray luminosity compared to the transient sources,
which makes the pulse-to-pulse technique probably the only way to
study the luminosity-dependence of the spectrum. Second, contrary to 
V0332+53 and 4U\,0115+63, a \emph{positive} correlation
of $E_{\rm cyc}$ and the X-ray luminosity was reported
on the basis of the entire set of \textsl{RXTE} observations of the source
\citep{Staubert:etal:07}. Using our method we were able to detect
a significant \emph{positive correlation} between $E_{\rm cyc}$ and the pulse
amplitude during a \emph{single} main-on state 
%covered by \textsl{RXTE} observations
where the averaged luminosity of the pulsar remained constant, thus,
confirming the finding of \citet{Staubert:etal:07}.
Additionally, we detected a negative correlation between the photon 
index $\Gamma$ and the pulse amplitude which is also contrary to 
V0332+53 and 4U\,0115+63 (see above), probably 
indicating a different accretion regime in this system (see Discussion
in \citealt{Staubert:etal:07}).
The photon index and the cyclotron line energy are shown in 
Fig.\,\ref{herx1frs} as a function of single pulse amplitude.

%%%%%%%%%%%%%% Figure 4 %%%%%%%%%%%%
%%%%%%%%%%%%%%%%%%%%%%%%%%%%%%%%%%%%

%- - - - - - - - - - - - - - - - - - - - - - - - - - - - - - - - - - 
%\subsection{A0535+26}

\begin{figure}
\centering
\begin{minipage}{0.45\textwidth}
  \epsfig{file=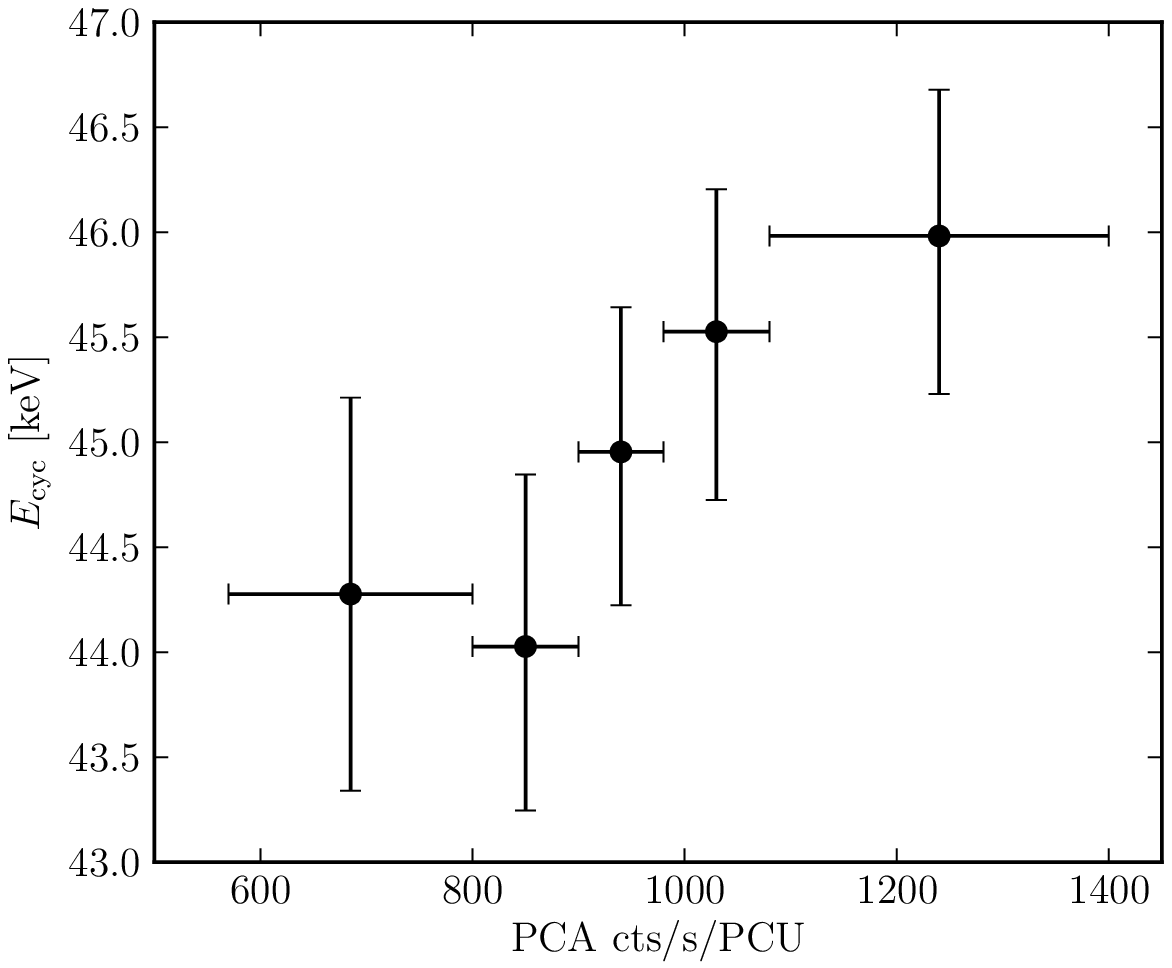, width=\textwidth}
\end{minipage}
\begin{minipage}{0.45\textwidth}
  \epsfig{file=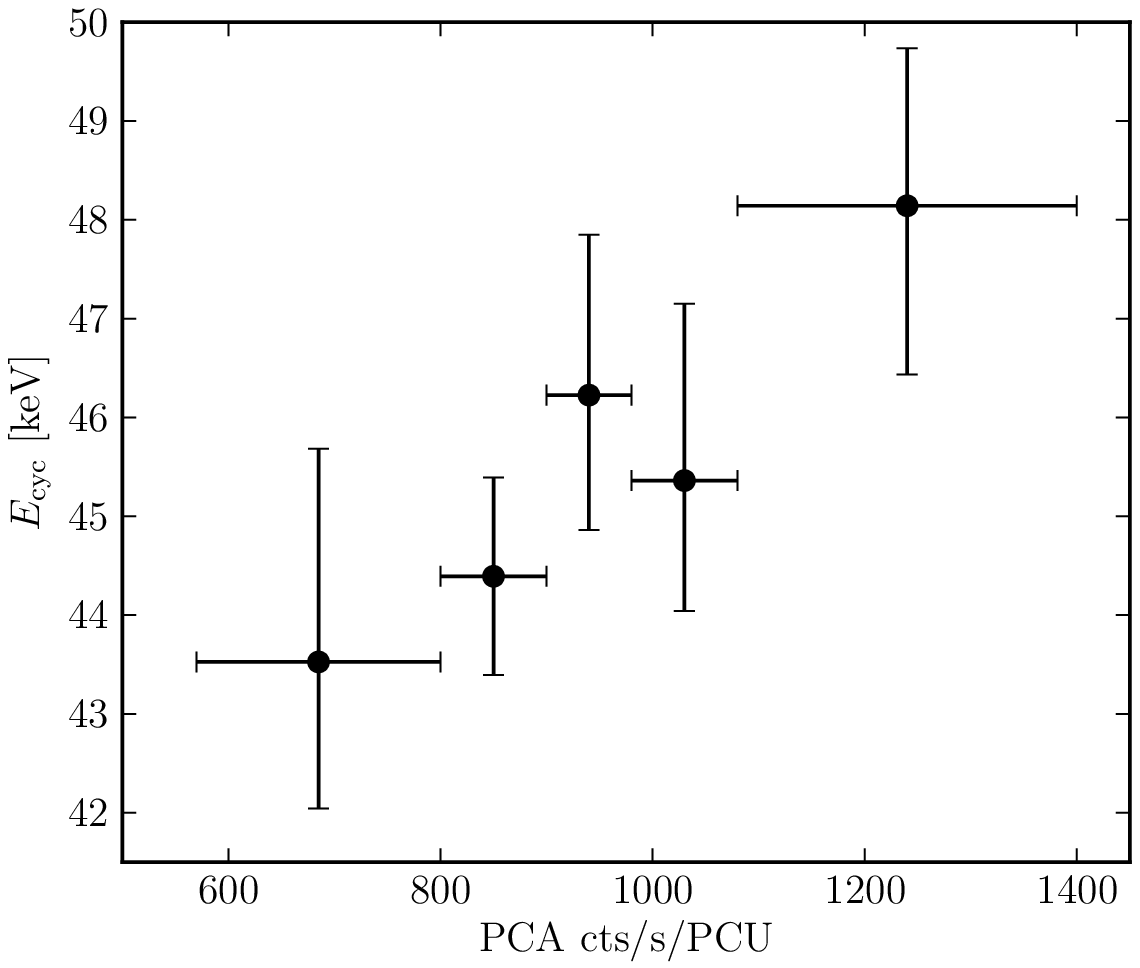, width=\textwidth}
\end{minipage}
\caption{The variation of the fundamental cyclotron line centroid energy 
  with the single pulse amplitude measured in A0535+26 with \textsl{RXTE} (left)
  and \textsl{Integral} (right).
  %The vertical bars indicate uncertainties at the 90\% c.l.
}
\label{a0535frs}
\end{figure}

\textbf{A0535+26.}
%Similarly to V0332+53 and 4U\,0115+63, A0535+26 is a transient
%Be/X-ray binary. However, 
Despite many observed outbursts,
no clear variations of the cyclotron line energy with flux have so far been 
reported \citep[][end references therein]{Caballero:etal:08}.
For this source our pulse-to-pulse spectroscopy %allowed to
revealed for the first time a \emph{positive} correlation between
the fundamental cyclotron line energy and the pulse
amplitude (Fig.\,\ref{a0535frs}). 
For A0535+26 we used the data obtained simultaneously
with \textsl{RXTE} and \textsl{INTEGRAL}. The \textsl{RXTE/PCA}
light curve was used to select %time intervals of 
pulses %with different amplitudes 
(for a subsequent spectral analysis) for both satellites. Therefore the X-axis
of the two graphs in Fig.\,\ref{a0535frs} (representing the pulse
amplitude) is in PCA count rate units. 
%(for both \textsl{RXTE} and\textsl{INTEGRAL} data). 
The correlation appears in the data
from both satellites which straightens the evidence. Our finding makes
A0535+26 only the second source (after Her~X-1) showing positive 
correlation of the cyclotron line energy with flux.

%%%%%%%%%%%%%% Figure 5 %%%%%%%%%%%%
%%%%%%%%%%%%%%%%%%%%%%%%%%%%%%%%%%%%

\section{Summary and conclusions}

The reported analysis of a sample of bright accreting 
pulsars allowed us for the first time to find
strong dependence of the X-ray spectral
continuum and the cyclotron line energy on the amplitude of individual
pulses. 
%Furthermore, the cyclotron line
%exhibited by all four pulsars was also found to vary significantly
%with the pulse height. 
We argue that the observed variability most probably reflects
the changing structure of the emitting region above the neutron star poles
which is able to adjust on a short time scale (seconds) 
to the variable accretion
rate $\dot M$. Indeed, the observed correlations of  $E_{\rm cyc}$
with the single pulse amplitude are in line with those 
reported on the basis of the averaged flux level for some of
the sources. The latter were interpreted assuming a variable height of the
accretion column which depends on $\dot M$
\citep[see discussions in][]{Mowlavi:etal:06,Staubert:etal:07}. 
Thus, similar changes of the polar accretion structure occurring 
on much smaller time scales are most
probably responsible for the pulse-to-pulse spectral variations
reported here.

\bibliographystyle{aa}
\bibliography{refs}

\end{document}